\documentclass[reprint,aps,prl,amsmath,amssymb]{revtex4-1}
\usepackage{color}
\usepackage{amsmath}
\usepackage{bm}
\usepackage{graphicx}
\usepackage{multirow}
\usepackage{float}

\begin{document}

\newcommand{\bn}{\bm n}
\newcommand{\bp}{\bm{p}}   
\newcommand{\br}{\bm{r}}
\newcommand{\bk}{\bm{k}}
\newcommand{\bv}{\bm{v}}
\newcommand{\brho}{{\bm{\rho}}}
\newcommand{\bj}{{\bm j}}
\newcommand{\wk}{\omega_{\bm{k}}}
\newcommand{\nk}{n_{\bm{k}}}
\newcommand{\eps}{\varepsilon}
\newcommand{\la}{\langle}
\newcommand{\ra}{\rangle}
\newcommand{\be}{\begin{eqnarray}}
\newcommand{\ee}{\end{eqnarray}}
\newcommand{\intl}{\int\limits_{-\infty}^{\infty}}
\newcommand{\dE}{\delta{\cal E}^{ext}}
\newcommand{\SE}{S_{\cal E}^{ext}}
\newcommand{\dsp}{\displaystyle}
\newcommand{\phit}{\varphi_{\tau}}
\newcommand{\p}{\varphi}
\newcommand{\cL}{{\cal L}}
\newcommand{\dphi}{\delta\varphi}
\newcommand{\dbj}{\delta\bm{j}}
\newcommand{\dI}{\delta I}
\newcommand{\dph}{\delta\varphi}
\newcommand{\ua}{\uparrow}
\newcommand{\da}{\downarrow}
\newcommand{\ip}{\{i_{+}\}}
\newcommand{\im}{\{i_{-}\}}
\newcommand{\tp}{\tilde p}
\newcommand{\change}[1]{{#1}}
\newcommand{\sgn}{{\rm sgn}\,}

\title{Two-Particle Scattering and Resistivity of Rashba Electron Gas}

\author{K. E. Nagaev} 


\affiliation{Kotelnikov Institute of Radioengineering and Electronics, Mokhovaya 11-7, Moscow 125009, Russia}


\begin{abstract}
We calculate the electrical resistivity  of a two-dimensional electron gas that results from two-particle collisions
and strong Rashba spin-orbit coupling. When combined with impurity scattering, the two-particle correction to the resistivity
is proportional to the square of temperature $T$ if only the lower helicity band is filled, but the $T^2$ term vanishes if the 
Fermi level is above the Dirac point. In the absence of impurities, two-particle collisions do not contribute to resistivity.
\end{abstract}

\maketitle

It is well known that electron-electron scattering does not affect the resistivity of Galilean-invariant Fermi liquids because
the current is proportional to the total momentum of electrons, which is conserved by interparticle collisions. However in 
realistic materials the electron-electron scattering may contribute to the resistivity through several mechanisms, which result 
in its $T^2$ temperature dependence. First of all, it is Umklapp scattering, which conserves the quasimomentum up to a reciprocal
lattice vector \cite{Peierls29,Landau36}. Baber \cite{Baber37} suggested that even normal collisions may result in the $T^2$ 
contribution to the resistivity of multi-band metals if the effective masses of electrons in the bands are essentially different. 
The combined action of electron--impurity and interband electron--electron scattering was considered in a large number of papers
both for two- (2D) and three-dimensional  systems \cite{Appel78,Murzin98,Hwang03}. Recently, this interplay was analyzed
for anisotropic Fermi surfaces \cite{Maslov11,Pal12-PRB,Pal12-LJP}. It was found that the $T^2$ term  is absent
for simply connected \change{convex} Fermi surfaces, but  is  present if they are concave or multiply connected.
Based on these findings, one may conclude that the existence of $T^2$ contribution to the resistivity from two-particle 
scattering for multiply connected Fermi surfaces depends only on their shape, but this is not the case. We demonstrate this 
using spin-orbit-couples 2D electron gas as an example.

In last decades, electric transport in 2D systems with Rashba spin-orbit coupling \cite{Bychkov84} became a subject of intensive 
investigations. This coupling
spin-splits the dispersion curve into the upper and lower helicity bands. It was found that in the low-density regime when only the
lower band is filled, the impurity-related resistivity exhibits an unconventional electron-density dependence 
\cite{Brosco16,Hutchinson18,Sablikov19}. 
Apparently, this system is not Galilean-invariant and is described at the same time by a minimum number of independent parameters. 
Therefore it is of interest to calculate its resistivity caused by two-particle collisions.

In this paper, we consider the effect of electron--electron scattering in a generic Rashba spin-orbit coupled 2D electron gas
\change{at temperatures smaller than the coupling energy}. 
Using the Boltzmann equation, we calculate the resistivity of this system both in the presence of impurities and in the pure case 
at high and low electron densities. Despite the doubly-connected Fermi surface, the contribution to the resistivity from the 
electon--electron collisions in the absence of impurities is zero.

\begin{figure}
 \includegraphics[width=1.0\columnwidth]{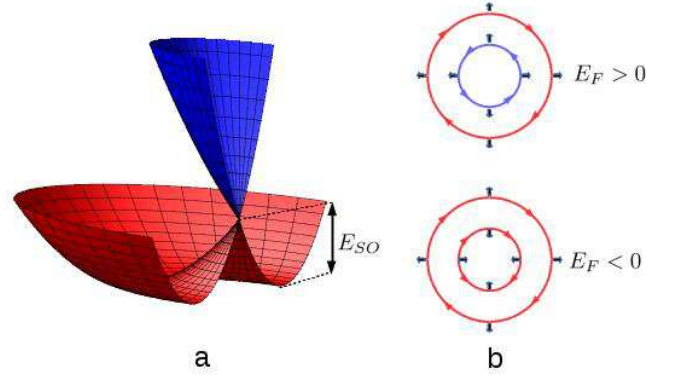}
 \caption{\label{fig:bands} (a) 3D plot of the lower (red) and upper (blue) helicity bands touching each other at the
 Dirac point. (b) The doubly connected Fermi surface above and below the Dirac point. Red and blue arrows show the directions
 of spin, black arrows show the directions of velocity at each contour of the surface. }
\end{figure}

Consider a two-dimensional electron gas lying in the $xy$ plane, so its unperturbed Hamiltonian is of the form
\be
 \hat{H} = \frac{\hat p_x^2 + \hat p_y^2}{2m} + \alpha\,(\hat\sigma_x\hat p_y - \hat\sigma_y\hat p_x),
 \label{H}
\ee
where $\alpha$ is the Rashba coupling constant and $\hat{\sigma}_{x,y}$ are the Pauli matrices. The diagonalization of this
Hamiltonian results in two branches of the spectrum
\be
 \eps_{\pm}(\bp) = \frac{p_x^2 + p_y^2}{2m} \pm \alpha\sqrt{p_x^2 + p_y^2},
\label{eps+-}
\ee
which correspond to the two rotationally symmetric nonparabolic energy bands with opposite helicities that intersect only in one
point at $p=0$ (the Dirac point, see Fig.~\ref{fig:bands}). The corresponding wave functions are spinors 
%
%
\be
 \Psi_{\bp\pm}(\br) 
 = \frac{1}{\sqrt{2}}\,e^{i\bp\br/\hbar} \binom{e^{i\chi_{\bp}/2}}{\pm e^{-i\chi_{\bp}/2}},
 \label{WF}
\ee
where $\chi_{\bp} = \arctan(p_x/p_y)$ so that the spin component perpendicular to $\bp$ is $\pm1/2$. The position of 
the Fermi level $E_F$ may be tuned by external electrostatic gates, so it can cross either only the lower band, or both bands.

The electron--impurity and electron--electron scattering is assumed to be weak, so the charge transport may be described
by the Boltzmann equation in the basis of eigenstates of Hamiltonian \eqref{H}. In the linear approximation in the
electric field $\bf E$, it is of the standard form \change{\cite{Haug-book}}
\be
 e{\bf E}\,\frac{\partial\eps_{\nu}}{\partial\bp}\,\frac{d \bar{f}}{d\eps_{\nu}} 
 = I_{\nu}^{imp}(\bp) + I_{\nu}^{ee}(\bp),
 \label{Boltz-lin}
\ee
where $\nu=\pm$ is the branch index and $\bar f$ is the equilibrium Fermi distribution. The electron--impurity collision 
integral may be written in the Born approximation as
\begin{multline}
 I_{\nu}^{imp}(\bp) = n_i \sum_{\nu'}\int \change{ \frac{d^2p'}{(2\pi\hbar)^2}  }\,
  \frac{2\pi}{\hbar}\,\left|U_{\bp\bp'}^{\nu\nu'}\right|^2\,\delta(\eps_{\nu}-\eps_{\nu'})
  \\ \times
  \left[f_{\nu'}(\bp') - f_{\nu}(\bp)\right],
 \label{Iimp}
\end{multline}
where $n_i$ is the concentration of impurities and $U_{\bp\bp'}^{\nu\nu'}$ is the matrix element of the impurity potential
between the electron states $(\bp,\nu)$ and $(\bp',\nu')$. In the case of point-like impurities with a potential $U(\br) = U_0\,\delta(\br)$, one easily obtains that
\be
 \left|U_{\bp\bp'}^{\nu\nu'}\right|^2 = \frac{1}{2}\,U_0^2\,
 [1 + \nu\,\nu' \cos(\widehat{\bp,\bp'})].
 \label{U^2}
\ee
The electron--electron collision integral is of the form
\begin{multline}
 I_{\nu}^{ee}(\bp) = \sum_{\nu_1} \sum_{\nu_2} \sum_{\nu_3}
 \int\frac{d^2p_1}{(2\pi\hbar)^2} \int\frac{d^2p_2}{(2\pi\hbar)^2} \int d^2p_3\,
 \\ \times
 \delta(\bp + \bp_1  -  \bp_2 - \bp_3)\,\delta(\eps_{\nu} + \eps_{\nu_1} - \eps_{\nu_2} - \eps_{\nu_3})
 \\ \times
 W^{\nu\nu_1,\nu_2\nu_3}_{\bp\bp_1, \bp_2 \bp_3} 
 \\ \times
 \bigl[ (1-f)(1-f_1)\,f_2\,f_3 - f\,f_1\,(1-f_2)(1 - f_3) \bigr].
 \label{Iee-gen}
\end{multline}
We assume that  due to the screening by a nearby gate, the interaction potential is short-ranged and may be written
as $V(\br-\br') = V_0\,\delta(\br-\br')$. Calculating the difference between the matrix elements of direct and exchange
interactions between the states \eqref{WF} and squaring it results in the expression for the scattering probability
\begin{multline}
 W^{\nu\nu_1,\nu_2\nu_3}_{\bp\bp_1, \bp_2 \bp_3} 
 = \frac{\pi}{2}\,\frac{V_0^2}{\hbar}\,
 [1 - \nu\,\nu_1 \cos(\widehat{\bp,\bp_1})]
 \\ \times
 [1 - \nu_2\,\nu_3 \cos(\widehat{\bp_2,\bp_3})].
 \label{W-expl}
\end{multline}
It is convenient to replace the momentum variable $\bp$ by the energy $\eps$ measured from $E_F$ and the angle $\p$ 
measured from the direction of electric field. The solution of Eq. \eqref{Boltz-lin} in band $\nu$ is sought in the standard form
\be 
 f_{\nu}(\eps,\p) = \bar f(\eps) + C_{\nu}(\eps)\,\bar f(\eps)\,[1 - \bar f(\eps)]\cos\p,
 \label{f-ansatz}
\ee
where $C_{\nu}$ describes the correction to $\bar f$ in the electric field.
A substitution of this ansatz into Eq. \eqref{Iee-gen} results in the linearized collision integral
\begin{multline}
 I_{\nu}^{ee}(\eps,\p) =  \sum_{\nu_1} \sum_{\nu_2} \sum_{\nu_3}
 \int \frac{d\eps_1}{|v_{\nu_1}|} \int \frac{d\eps_2}{|v_{\nu_2}|} \int \frac{d\eps_3}{|v_{\nu_3}|}
 \\ \times 
 \delta(\eps + \eps_1 - \eps_2 - \eps_3)\,(1 - \bar{f})(1 - \bar{f}_1)\,\bar{f}_2\,\bar{f}_3
 \\ \times
 \bigl( \Omega_2\,C_{\nu_2} + \Omega_3\,C_{\nu_3} - \Omega_1\,C_{\nu_1} - \Omega\,C_{\nu} \bigr)
 \cos\p,
\label{Iee-lin}
\end{multline}%
where the quantities
\begin{multline}
 \Omega_i(\eps...\eps_3,\nu...\nu_3)
 = 
 \int\frac{d^2p_1}{(2\pi\hbar)^2} \int\frac{d^2p_2}{(2\pi\hbar)^2} \int d^2p_3
 \\ \times
 W^{\nu\nu_1,\nu_2\nu_3}_{\bp\bp_1, \bp_2 \bp_3}\,
 \cos(\widehat{\bp,\bp_i})\,
 \delta(\bp + \bp_1  -  \bp_2 - \bp_3)
 \\ \times
 \delta(|\bp_1|-p_{\nu_1})\,\delta(|\bp_2|-p_{\nu_2})\,\delta(|\bp_3|-p_{\nu_3})\,
 \label{Omega-gen}
\end{multline}
include both the scattering parameters and the effective phase volume available for the scattering,
$p_{\nu_i}(\eps_i)$ is the solution of equation $\eps_{\nu_i}(p)=E_F+\eps_i$, and $v_{\nu_i} = (dp_{\nu_i}/d\eps_i)^{-1}$ is 
the corresponding velocity.

The interband scattering is relevant only if the Fermi level crosses both helicity bands. However if it is located below 
the Dirac point, the Fermi surface is still doubly connected because of nonmonotonic $\eps_{-}(p)$ dependence. In this case,
one can use the above equations by replacing labels $\nu_{\pm}$ with $\lambda = \gtrless$ that correspond to the ascending 
and descending 
portions of this curve. The products $\nu\nu_1$ and $\nu_2\nu_3$ must be set equal to 1 and the remaining $\nu_i$ must be
replaced by $\lambda_i$. However it should be kept in mind that $v_{>}(\eps) = -v_{<}(\eps)$, while $v_{+}(\eps)=v_{-}(\eps)$.
Depending on the sign of $E_F$, we denote either $\nu$ or $\lambda$ by $\mu$ where it does not lead to a confusion and imply 
that $-\mu$ reverses the sign of $\nu$ or sense of $\lambda$. For example, the equation for the current density may be
written as 
\be
 j = \frac{e}{4\pi\hbar^2} \sum_{\mu} \sgn v_{\mu} \int d\eps\,{\bar f}(1 - \bar f)\,  p_{\mu}\,C_{\mu}.
 \label{j-gen}
\ee

First of all we calculate the collision integrals $I_{\mu}^{imp}$ and $I_{\mu}^{ee}$ assuming  
$C_{\mu}$ and all other quantities except the distribution functions $\bar f_i$ to be energy-independent. 
The electron--impurity collision integrals are easily calculated and equal
\begin{multline}
 I_{\mu}^{imp}(\eps,\p) = 
 -\Gamma_0\, \cos\p\,{\bar f}(1 - \bar f)
 \bigl[ (p_{\mu} + 2\,p_{-\mu})\,C_{\mu} 
 \\+
 \sgn E_F\, p_{-\mu}\,C_{-\mu} \bigr]
 /(p_{\mu} + p_{-\mu}),
 \label{Iimp-ab}
\end{multline}
where $\Gamma_0 = n_i U_0^2\,(p_{\mu} + p_{-\mu})/4\hbar^3 |v_{\mu}|$ and
all quantities except $\bar f$ are taken at $\eps=0$. The sign between the terms in the square brackets
depends on whether the electron velocity at the inner and outer Fermi contours has the same or opposite signs.
If only the impurity scattering is present, the system of equations \eqref{Boltz-lin} is easily solved in $C_{\mu}$ 
to give
\be
 C^{imp}_{\mu} = 
 eE\,v_{\mu}\,(\Gamma_0\, T)^{-1}\,p_{\mu}/(p_{\mu}+p_{-\mu}),
 \label{C-el}
\ee
and calculating the current via \eqref{j-gen} gives the same current density as in Refs. \cite{Brosco16,Hutchinson18}.

\begin{figure}
 \includegraphics[width=0.9\columnwidth]{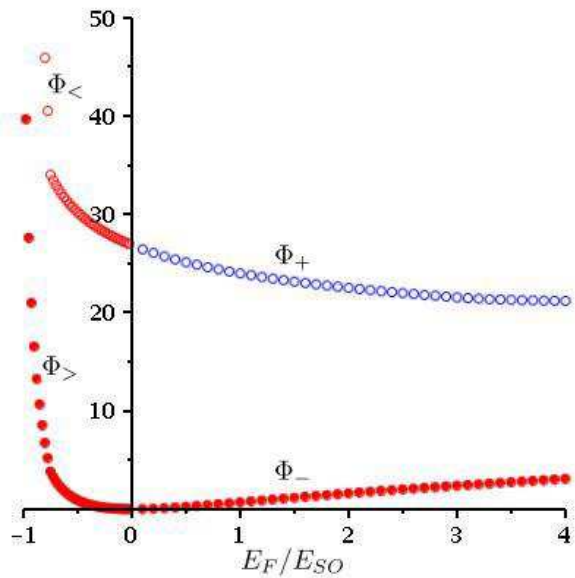}
 \caption{\label{fig:Phi} The dependences of $\Phi_{\mu}$ on $E_F/E_{SO}$. $\Phi_{\gtrless}$
 exhibit a logarithmic singularity at $E_F=-E_{SO}$, and $\Phi_{\pm}(\infty)=16$. }
\end{figure}

Under the same conditions, the collision integral \eqref{Iee-lin} is proportional to $T^2$ and may be brought to the form
\begin{multline}
 \bar I_{\mu}^{ee}(\eps,\p) = 
 \Gamma_2(T)\,\cos\p\,\bar{f}(1-\bar{f})\,(\eps^2/T^2+\pi^2)
 \\ \times
 \Phi_{\mu}(E_F/E_{SO})\,(p_{\mu}C_{-\mu} - p_{-\mu}C_{\mu})/(p_{\mu}+p_{-\mu}),
 \label{Iee-2}
\end{multline}
where $\Gamma_2(T)=V_0^2 T^2\,(p_{\mu}+p_{-\mu})/32\pi^3\hbar^5|v_{\mu}|^3$
and $E_{SO}=\alpha^2/2m$ is the characteristic energy of spin-orbit coupling.  The dimensionless functions 
$\Phi_{\mu}$ \change{represent the phase volume available for the scattering} and can be calculated only numerically 
\change{(see Fig. \ref{fig:Phi} and Appendix for details)}. 
Except for the explicit form
of $\Gamma_2$ and $\Phi_{\mu}$, Eq.~\eqref{Iee-2} does not depend on the presence of spin-orbit coupling.
It is clearly seen that in this 
approximation, $I_{\mu}^{ee}$ is zero for any distribution function of the form \eqref{f-ansatz} with 
$p_{\mu}C_{-\mu} = p_{-\mu}C_{\mu}$. Therefore in the absence of impurity scattering, the system of kinetic equations 
\eqref{Boltz-lin} with $I_{\mu}^{ee}$ given by Eq. \eqref{Iee-2} becomes degenerate and has no stationary solution. 

To overcome this difficulty, we consider the case where the impurity scattering is strong and $I^{ee}$ may be treated 
as a perturbation. The solution of Eq. \eqref{Boltz-lin} is sought as a sum $f = \bar{f} + \delta f^{imp} + \delta f^{ee}$,
where $\delta f^{imp}$ is given by Eq. \eqref{f-ansatz} with $C^{imp}_{\mu}$ from Eq. \eqref{C-el}, and $\delta f^{ee}$ is the 
solution of equation 
\be 
 I_{\mu}^{imp}\{\delta f^{ee}\} + \bar I_{\mu}^{ee}\{\delta f^{imp}\} = 0.
 \label{f1-eq}
\ee
This equation is easily solved for $C^{ee}_{\mu}$ and the correction to the current is calculated using Eq. \eqref{j-gen}.
Below the Dirac point, it equals
\begin{multline}
 \delta j^{ee}
 =
 -\change{
 \frac{2\pi e^2E}{3\,\hbar^2}\,}
 \frac{\Gamma_2}{\Gamma_0^2}\,v_{>}\,
 \frac{p_{>}p_{<} \left[ p_{<}^2\,\Phi_{<} + p_{>}^2\,\Phi_{>}\right]}
 {(p_{>}+p_{<})^3}.
 \label{j-imp+ee}
\end{multline}
The inelastic correction to the current \change{ tends to zero as $E_F \to 0$  and vanishes above $E_F$} 
because $C_{\gtrless}$ from 
Eq.~\eqref{C-el} turn $I_{\gtrless}^{ee}\{\delta f^{imp}\}$ into zero. Actually this is a consequence of equal slopes of
the two dispersion curves $\eps_{+}(p)$ and $\eps_{-}(p)$ at the same energy. \change{The correction logarithmically 
diverges at the bottom of the lower helicity band because of singularity in $\Phi_{\gtrless}$ due to head-on collisions
of electrons on the different Fermi contours as $p_{>}$ and $p_{<}$ approach each other. This implies that the
perturbative result  Eq. \eqref{j-imp+ee} breaks down in this limit. }

The contribution to the resistivity from electron--electron scattering  is larger than that from electron--phonon scattering,
which is proportional to $T^{4.5}$ at low temperatures \cite{Kawamura92}.
The possibility of observing  it depends on the quality of the samples. A good 
candidate for such experiments is 2D electron gas in InAs, which exhibits a strong spin-orbit coupling with Rashba parameter
$\hbar\alpha =1.2\,{\rm eV\AA}$ \cite{Heedt2017}. The electron-electron scattering effects are more prominent at low
concentrations when only the lower helicity band is filled.  For the electron concentration $3\cdot 10^{10}$ cm$^{-2}$, 
the gas--gate distance of 20 nm, and for the elastic mean free path of 800 nm reported very recently in InAs 2D electron gas in Ref. \cite{Lee19}, the  temperature-dependent correction to the resistivity may be of the same order as the impurity-induced
resistivity already at $T= 2$ K. Therefore it may be observable for realistic parameters of the system.



\begin{acknowledgements}
This work was supported by Russian Science Foundation (Grant No 16-12-10335).
\end{acknowledgements}

\bibliographystyle{my_pss}
\bibliography{paper}

\appendix
\begin{widetext}
\section{Expressions for $\Phi_{\gtrless}$}

The quantities $\Phi_{\gtrless}$ that appear in the electron--electron collision integral  Eq. (15) below 
the Dirac point are obtained as the sums of integrals
\be
 \Phi_{<} =  \sum_{i=1}^8 \zeta_i(x),
 \quad
 \Psi_{>} =  \sum_{i=1}^{8} \zeta_i(1/x),
\ee
where
\begin{subequations}
\label{zeta-i}
\begin{gather}
 \zeta_1(x) = 0,
 \\
 \zeta_2(x) =  \zeta_3(x) = \int\limits_{-\pi}^{\pi} d\p\,(1-\cos\p)\,
 \frac{1+x^2+2\cos\p}{4\,x}\,
 \left|{\rm Re}\,\sqrt{\frac{(1+x)^2/2-1-\cos\p}{1+\cos\p-(x-1)^2/2}}\right|,
 \\
 \zeta_4(x) = \frac{1}{x}\int\limits_{-\pi}^{\pi} d\p\,(1-\cos\p)\,
 \left|{\rm Re}\,\sqrt{(2\,x^2-1-\cos\p)(1+\cos\p)}\right|,
 \\
 \zeta_5(x) = -x \int\limits_{-\pi}^{\pi} d\p\,(1-\cos\p)\,\cos\p\,
 \left|{\rm Re}\,\sqrt{\frac{3/x-x-2\cos\p}{x+1/x+2\cos\p}}\right|,
 \\
 \zeta_6(x) = \zeta_7(x) = x\int\limits_{-\pi}^{\pi} d\p\,(1-\cos\p)^2\,\frac{\sqrt{1-\cos^2\p}}{x+1/x+2\cos\p},
 \\
 \zeta_8(x) = \int\limits_{-\pi}^{\pi} d\p\,(1-\cos\p)\,
 \left|{\rm Re}\,\sqrt{\frac{3\,x-1/x-2\cos\p}{x+1/x+2\cos\p}}\right|,
\end{gather}
\end{subequations}
and $x=p_{>}/p_{<}$. The transition to the dependence on $E_F$ below the Dirac point is performed by means of equation
\be
 x = \frac{1 + \sqrt{1 + E_F/E_{SO}}}{1 - \sqrt{1 + E_F/E_{SO}}}.
 \label{x}
\ee

Quantities $\Phi_{\pm}$ that appear in the electron--electron collision integral  Eq. (15) above the Dirac point
are represented by the sums of expressions
\be
 \Phi_{-} =  \sum_{i=1}^8 \xi_i(y),
 \quad
 \Phi_{+} =  \sum_{i=1}^8 \xi_i(1/y),
\ee
where
\begin{subequations}
\label{xi-i}
\begin{gather}
\xi_1(x)=0,
\\
\xi_2(y)= \xi_3(y) = \int\limits_{-\pi}^{\pi} d\p\,
\left( 1 -\cos \p \right) \frac{ 2\,{y}^{2}\cos \p +{y}^{2}+1 }{4\,y}\,
\left|{\rm Re}\,\sqrt{\frac{(y^{-1}-1)^2/2 - 1 - \cos\p}{1 + \cos\p - (y^{-1}+1)^2/2}}\right|,
\\
\xi_4(y)= \int\limits_{-\pi}^{\pi} d\p\,
\left(1 - \cos \p \right)  
\sqrt{\cos \p +1} \left|{\rm Re}\,\sqrt{2 -{y}^{2}\,(1 +\cos \p)}\right|,
\\
\xi_5(y)=- \int\limits_{-\pi}^{\pi} d\p\,
\frac{\cos\p \left( \cos \p +1 \right)}{y} 
\left|{\rm Re}\,\sqrt{{\frac { 3y - 1/y - 2\cos \p }{y + 1/y + 2\cos \p}}}\right|,
\\
\xi_6(y)= \xi_7(y)= \int\limits_{-\pi}^{\pi} d\p\,
{\frac { \left( \cos \p +1 \right) ^{2} 
\sqrt{1 - \cos^2 \p}}{{y}^{2} + 2y \cos \p +1}},
\\
\xi_8(y)= \int\limits_{-\pi}^{\pi} d\p\,
\left( \cos \p +1 \right)  
\left|{\rm Re}\,\sqrt{{\frac {3/y - y -2\,\cos \p}{y + 1/y + 2\,\cos \p}}}\right|,
\end{gather}
\end{subequations}
and $y= p_{-}/p_{+}$. The transition to the dependence on $E_F$ above the Dirac point is performed by means of equation
\be
 y = \frac{\sqrt{E_F/E_{SO}+1} + 1}{\sqrt{E_F/E_{SO}+1} - 1}.
 \label{y}
\ee
Overall dependences of quantities $\Phi_{\gtrless}$ and $\Phi_{\pm}$ on $E_F$ are shown in 
Fig. 2 of the paper.
At $E_F= -E_{SO}$, both $\Phi_{<}$ and $\Phi_{>}$ exhibit a logarithmic singularity
\be
 \Phi_{\gtrless}(E_F/E_{SO}) = 16\,\ln\!\left(\frac{E_{SO}}{E_F+E_{SO}}\right).
 \label{EF=-ESO}
\ee
At the Dirac point $E_F=0$, $\Phi_{\gtrless}$ smoothly join $\Phi_{\pm}$, so that
\be
 \Phi_{<}(0) = \Phi_{+}(0) = 16 + 2\sqrt{3}\pi,
 \quad
 \Phi_{>}(0) = \Phi_{-}(0) =0.
 \label{EF=0}
\ee
In the limit $E_F/E_{SO} \to\infty$, both $\Phi_{+}$ and $\Phi_{-}$ tend to the same limiting value
\be
 \Phi_{+}(\infty) = \Phi_{-}(\infty) = 16.
 \label{EF=inf}
\ee
\end{widetext}

\end{document}